\documentclass[12pt]{article}
\usepackage{putex}
\usepackage{graphicx}

\begin{document}

\preprint{hep-th/0402225 \\ PUPT-2112}
\institution{PU}{Joseph Henry Laboratories, Princeton University, Princeton, NJ 08544}

\title{Structure formation in a string-inspired modification of the cold dark matter model}

\authors{Steven S. Gubser\footnote{e-mail: \tt ssgubser@Princeton.EDU} and P.J.E. Peebles\footnote{e-mail: \tt pjep@Princeton.EDU}}

\abstract{Motivated in part by string theory, we consider the idea that the standard $\Lambda$CDM cosmological model might be modified by the effect of a long-range scalar dark matter interaction. The variant of this widely-discussed notion considered here is suggested by the Brandenberger-Vafa \cite{bv} picture for why we perceive three spatial dimensions. In this picture there may be at least two species of dark matter particles, with scalar ``charges'' such that the scalar interaction attracts particles with like sign and repels unlike signs. The net charge vanishes. Under this condition the evolution of the mass distribution in linear perturbation theory is the same as in the $\Lambda$CDM cosmology, and both models therefore can equally well pass the available cosmological tests. The physics can be very different on small scales, however: if the scalar interaction has the strength suggested by simple versions of the string scenario, nonlinear mass concentrations are unstable against separation into charged halos with properties unlike the standard model prediction and possibly of observational interest.}

\date{February, 2004}
\maketitle

\section{Introduction}
\label{INTRODUCTION}

In this paper and its companion \cite{gpTwo} we investigate models for a  long-range scalar-mediated interaction among dark matter particles.  Our starting assumptions are that the scalar in question is massless, or very nearly so, and that the scalar has negligible effect on physics in the visible sector---apart from what results from the gravitational interaction with the dark sector.  The fundamental theory that motivates our models offers reasons to doubt both assumptions, but the arguments are not convincing enough to discourage investigation of the possible observational significance of a string-inspired scalar field modification of the cold dark matter (CDM) cosmological model.

In addition to ideas from superstring theory, we are motivated by the thought that there surely is considerable room in the dark sector of cosmology for richer physics than the standard CDM model. Many current notions about viable examples of richer physics trace back to the long history of ideas about scalar-tensor gravity physics, elements of which are reviewed in a paper by G. Farrar and one of us \cite{fp}.\footnote{We take this opportunity to remedy an omission: the survey should have included the work on this subject by Frieman and Gradwohl \cite{Frieman:1993fv}.} The convergence of these long-standing ideas about gravity physics with what superstring theory might be suggesting is appealing and just might signify that we are being led to a better approximation to cosmology. 

The inspiration from string theory is the Brandenberger-Vafa \cite{bv} picture for why we perceive three spatial dimensions. It suggests that the long-range interaction of dark matter is the sum of the gravitational force and the force from the exchange of massless scalars and gauge bosons. The dark matter particles may be assigned ``charges'' ---to be defined more precisely in section~\ref{BV}---as in electrostatics, but particles with like signs of the scalar charge are attracted and unlike scalar signs are repelled. The model requires zero net scalar charge, which implies that in linear perturbation theory the evolution of the mass disttribution is unaffected by the scalar interaction. This means that initial conditions can be chosen so our model and the $\Lambda$CDM cosmology equally well pass the cosmological tests, including the new tests based on the measurements  of the large-scale distributions of mass and the 3~K thermal background radiation \cite{Bennettetal}. The scalar interaction introduces a new mode of instability, however, in which particles with different signs of charge tend to concentrate in different places. If there were a significant primeval amplitude in this charge separation mode it would tend to produce observable mass concentrations that are not closely correlated with the large-scale mass distribution, an observationally unacceptable situation. This is avoided if the initial conditions are very close to adiabatic, as we will assume. If the scalar force were comparable to the prediction of the simplest of string theory setups, then even with purely adiabatic initial conditions a strongly nonlinear dark matter mass concentration would be unstable against separating into massive charged halos, and halos of different signs would tend to avoid each other. We will suggest that such curious behavior may have something to do with the confusing situation in the theory and phenomenology of structure formation on the scale of galaxies. Other viable models for a modified CDM cosmology would have different implications for small-scale structure, of course. An example based on a supersymmetric picture for the dark matter, which allows a non-zero net scalar charge, is discussed in our companion  paper \cite{gpTwo}.

We organize this paper as follows.  In section~\ref{BV} we introduce the Brandenberger-Vafa scenario and describe its late-time dynamics in terms of an effective lagrangian for the dark matter.\footnote{It should perhaps be emphasized that, for us, late time means the matter- and radiation-dominated epochs: this makes our investigation somewhat different in spirit from others in the brane gas literature.}  In section~\ref{PERTURBATIONS}, we characterize the long-range forces in the dark sector by a matrix of dimensionless quantities which encode the physics of subsequent interest at scales comparable to the sizes of galaxies.  Using this matrix, we demonstrate in examples how the scalar force leads to a charge separation instability in the mass distribution. In section~\ref{OBSERVATIONS}, we discuss possible observational consequences of this effect for structure formation on the scale of galaxies, and we comment on some issues of small-scale extragalactic structure that might point to physics along the lines of this model. These considerations are continued in our companion paper \cite{gpTwo}.  In the last section, we return to the fundamental issues mentioned in our introductory paragraph: we consider why the scalar force does not manifest itself in the visible sector, in the E\"otv\"os experiment and measurements of orbits in the Solar System; we discuss the naturalness of assuming such a small mass for the scalar; and we outline some other issues of the fundamental physics of the model.

\section{The Brandenberger-Vafa scenario}
\label{BV}

Brandenberger and Vafa \cite{bv} put forward a simple and intriguing rationale, based on string theory, for space to appear three-dimensional.  Suppose, the argument goes, the universe started off very small and very hot, with all of the nine spatial dimensions predicted by string theory curled up at the string scale (which is close to the Planck scale), and all the degrees of freedom of string theory thermally excited, including strings wrapping all possible directions and strings with momenta in all possible dimensions.  If three spatial dimensions start to grow large, then the strings wrapping these dimensions will tend to annihilate efficiently with one another, because two extended strings moving in three spatial dimensions will generically intersect at some time.  But if four or more dimensions start to grow large, the strings wrapping them will not generically intersect, so they will not annihilate one another efficiently, and their tension will draw the size of the dimensions back toward the initial state.  At late times, then, it makes sense for three dimensions to have grown large, the strings extended in these dimensions to have largely annihilated, and the remaining six dimensions to have stayed small.

The strings wrapping and winding the six extra dimensions may stabilize the size of those dimensions: winding strings are lighter when the extra dimensions are smaller, but momentum modes are lighter when the extra dimensions are larger; thus there is a balance point depending on the relative numbers of winding and momentum modes.  A recent numerical study of this was presented in \cite{bw}, for perturbations with no spatial dependence. The basic idea of stabilizing moduli of string theory with winding states actually predates \cite{bv}: the earliest reference we know is \cite{kp}.

The starting point for the action for dark matter particles in the late-time picture inspired by the Brandenberger-Vafa scenario (as in \cite{bw}), for many discussions of scalar-tensor gravity models, and for our investigation, is
 \eqn{Action}{
  S = \int d^4 x \sqrt{g} \left[ {R \over 16\pi G} - 
   {1 \over 2} (\partial\phi)^2 - V(\phi) \right] - 
   \sum_\alpha \int_{\gamma_\alpha} ds \, m_\alpha(\phi) \,.
 }
The sum is over all particles in the dark sector. For simplicity we ignore the visible sector of baryons,  electromagnetic radiation  and neutrinos; the focus of our discussion is the growth of cosmic structure at relatively low redshifts, when the dark matter dominates. We eventually take the potential $V(\phi)$ to be a constant, but it is instructive to keep the general function for now. For simplicity we will also ignore the cosmological constant (or dark energy), that is, we eventually set $V(\phi) = 0$.  At late time in the Brandenberger-Vafa scenario, the winding and momentum modes have shed any excitations they carried, and they are heavy (on order the string scale in conventional setups) and their interactions are feeble, so it does not matter whether they are fermions or bosons, and the world-line description used in \Action\ is appropriate \cite{fp}.

Neglecting spacetime curvature fluctuations, the wave equation for the scalar field is
 \eqn{Waveequation}{
 \square\phi - dV/d\phi - \sum_q n_qdm_q/d\phi =0\, ,
 }
where $n_q$ is the number density of particles of species q and we are assuming the particle peculiar velocities are nonrelativistic. The time-scale for relaxation of the value of $\phi$ to the minimum of its potential may be comparable to the Hubble time. If $\phi$ is still rolling to its minimum it can cause serious problems with the cosmological tests \cite{fp}. We assume therefore that $\phi$ has relaxed close to its minimum or is placed there as an initial condition. 

On scales small compared to the Hubble length, and ignoring  $V(\phi)$, one sees from \Waveequation\ that the force between two dark matter particles is
 \eqn{GravityScalar}{
  F = {G m_1 m_2 \over r^2} + 
   {dm_1 \over d\phi} {dm_2 \over d\phi} {1 \over 4\pi r^2} \,.
 }
The second term comes from the scalar exchange, computed in linear perturbation theory. The $dm_q/d\phi$ thus are evaluated at the mean field value.  Because of the analogy with the Coulomb force law, we will term $Q_q = dm_q/d\phi$ the scalar charge of particle $q$.  Scalar forces are different from the gauge forces familiar from electrostatics: particles with like scalar charge sign attract and pairs with opposite sign are repelled.\footnote{It is worth bearing in mind that scalar charge as we have defined it is not actually a conserved charge.  For example, a heavy particle and its anti-particle have the same scalar charge because their mass is the same for every value of $\phi$, so total scalar charge can change via pair annihilation.  And if particles with scalar charge form a black hole, the scalar charge is lost.}  Signs are chosen so that $F$ in \GravityScalar\ is positive when the total force is attractive.

As we will describe more detail in section~\ref{CHARGES}, the scalar charge-to-mass ratio $Q_q/m_q$ is expected to be on the order of $1/M_{\rm Pl}$, where by definition $M_{\rm Pl}^{-2} = 8\pi G$.  So the scalar force between particles can be expected to be comparable to the gravitational term, independent of separation. We see from \Waveequation,
however, that when $V(\phi)$ is negligible and $\phi$ has relaxed to the minimum of its energy the mean scalar charge density vanishes. This means that there is no long-range scalar force when the space distributions of the dark matter particles are all the same. Thus we can construct viable models in which the scalar force affects the net mass distribution only where nonlinear dynamics has driven separation of the scalar charges. 

\section{Density perturbations}
\label{PERTURBATIONS}

We consider here in linear perturbation theory the behavior of inhomogeneous distributions of the dark matter components in an expanding universe. The first subsection generalizes the usual theory for pressureless freely moving  matter to take account of scalar and gauge charge interactions among the dark matter components. In section~\ref{SOLUTIONS} we consider the nature of solutions to the perturbation equations. The string theory considerations in section~\ref{CHARGES} lead us to our working assumption, as the simplest possibility, that there are two dark matter components with masses close to the four-dimensional Planck mass, scalar charges near $\pm 1$, and  gauge charges at a similar scale which however do not substantially affect the dynamics.

\subsection{Linear Perturbation description of the Scalar Interaction}
\label{LINEAR}

One can start from the action in equation \Action , with a specified functional dependence of the masses of the winding and momentum modes on the scalar field value, introduce a hydrodynamic approximation, and work out the full set of controlling perturbation equations. Anticipating that some readers may be unfamiliar with density perturbation calculations in cosmology, we do this in Appendix~A, arriving at a special case of equation \eno{NewDM} below.  The derivation presented here follows by inspection of the usual analysis of the gravitational growth of structure, which we need only modify by adding to the gravitational acceleration the acceleration caused by the scalar field inverse square force law. 

We will write the net force between particles $q$ and $p$ as
\eqn{BetaDefine}{
  F_{pq} = \beta_{pq} {G m_p m_q \over r^2} \,, \qquad
  \beta_{pq} = 1 + {Q_pQ_q\over 4\pi G m_pm_q}\,,
  }
for $N$ different species of dark matter particles. As in equation~\GravityScalar , the scalar charge is $Q_q = dm_q/d\phi$, and the condition that the field is locked by its effective potential $V_m = \sum n_qm_q$ is that the total scalar charge vanishes, 
\eqn{zerocharge}
{\sum_qn_qQ_q = 0\,,} 
where $n_q$ is the mean number density. These equations give the full nonlinear description of the dynamics of structure formation on scales small compared to the Hubble length (and assuming the dark matter particle velocities are nonrelativistic, which is realistic well away from black holes, and ignoring baryonic matter, which a useful first approximation). They also give an easy guide to the linear perturbation equations for the evolution of the distributions of the dark matter components, as follows. 

Recall that when pressure and all other nongravitational forces may be ignored the mass density contrast, $\delta_m ({\bf x},t) = \delta\rho ({\bf x},t)/\rho (t)$, satisfies the linear perturbation equation 
 \eqn{KeyCDM}{
  \ddot\delta_m + {2\dot{a} \over a} \dot\delta_m =
    4\pi G \rho \delta_m \,,
 }
where $\rho (t)$ is the total mean mass density, $a$ is the scale factor, and dots denote derivatives with respect to proper cosmological time.  The second term arises from the decay of peculiar velocities caused by the expansion of the universe, and the right hand-side represents the driving term due to the acceleration of gravity.  If there are several distinct species of cold dark matter particles, still with only gravitational interactions, then the evolution of each  density contrast $\delta_q$ is driven by the same gravitational acceleration. The perturbation equations thus are
 \eqn{SeveralCDM}{
  \ddot\delta_q + {2\dot{a} \over a} \dot\delta_q = 
    4\pi G \rho \delta_m = 4\pi G \sum_p m_pn_p\delta_p \,.
 }
The mean mass density in dark matter of type $p$ is $\rho_p = n_pm_p$, and the total density contrast is 
\eqn{deltam}{
\delta_m = \sum {\rho_p\delta_p/\rho } = \sum f_p\delta_p,
}
where $f_p$ is the mass fraction in dark matter of type $p$: $f_p = n_p m_p / \sum_q n_q m_q$.

The scalar interaction  changes \SeveralCDM\  by adding to the right-hand side of the equation the term representing the acceleration of matter of type $q$ by the scalar interaction with matter of type $p$: 
 \eqn{NewDMa}{
  \ddot\delta_q + {2\dot{a} \over a} \dot\delta_q =
    4\pi G \sum_p n_p\delta_p 
     \left( m_p + Q_pQ_q/ 4\pi G m_q\right) \,.
 }
The ratio of the two terms in parentheses is defined in equation~\BetaDefine , so we can rewrite this expression as
\eqn{NewDM}{
  \ddot\delta_q + {2\dot{a} \over a} \dot\delta_q =
    4\pi G \rho \sum_p \beta_{qp} f_p \delta_p \,.
 }
This is the desired result.

It is an easy exercise to obtain equation~\NewDM\  in a more direct way from the linearized expressions for mass conservation and the equation of motion, along with Poisson's equations for the gravitational and scalar forces, as in \cite{PeeblesFirstBook}. Yet another approach to this result is presented in Appendix~A.

The dark matter components may have gauge charges $Q^{\rm g}_p$ as well as scalar charges $Q^{\rm s}_p$. Since peculiar velocities are small, and the gauge charge-to-mass ratio suggested by string theory is exceedingly small compared to electromagnetism in the visible sector, the gauge interaction simply adds another long-range force that brings the matrix in equation~\BetaDefine\ to
 \eqn{NewBeta}{
  \beta_{pq} = 1 + {\sum_s Q^s_pQ^s_q - 
   \sum_{\rm g} Q^{\rm g}_pQ^{\rm g}_q\over 4\pi Gm_pm_q} \,,
 }
where we have summed over ${\rm s}$ and ${\rm g}$ to indicate that there may be several different scalar or gauge interactions.  Under each type of interaction, the net charge must vanish.  In the case of gauge charges, this follows from Gauss's Law, while for the various scalars, the argument again is that the effective potential for the scalars is minimized.  Thus
 \eqn[c]{NetChargeVanishes}{
\sum_p f_p Q^{\rm s}_p/m_p = 0, \qquad \sum_p f_p Q^{\rm g}_p/m_p = 0, 
 \cr\sum_p f_p \beta_{pq} = 1 \,,
 }
where the second line follows from the first applied to \NewBeta.

In the following subsection, we will exhibit solutions to \NewDM\ in the regime where non-relativistic matter is the dominant component of the energy density of the universe.  Let us give here a brief intuitive preview of the results and comment briefly on the evolution of perturbations in earlier epochs.

First, there is always an adiabatic mode whose evolution in the linear regime is exactly as it would be in the standard $\Lambda$CDM model.  To show this, multiply \NewDM\ by $f_q = m_q n_q/\rho$ and sum over $q$.  Second, there are isocurvature charge separation modes: a gauge charge or a scalar charge (or both) develops locally, compensated by an opposite charge elsewhere.  In the case of gauge charges, such charge separation grow less fast than the adiabatic mode.  This is because in the absence of gravity, a charged plasma dynamically seeks local charge neutrality and leads to Debye screening for the photon.  Exactly the opposite is true of scalar charges because like charges attract and opposite charges repel: isocurvature modes which lead to separation of scalar charges can grow faster than the adiabatic mode.  We will see that a reasonable suppression of isocurvature modes in the initial spectrum of perturbations avoids conflict with observation from the linear regime.  In section~\ref{OBSERVATIONS} we will consider the non-linear regime, where scalar charge separation modes are important even when isocurvature modes are suppressed in the initial spectrum.

We remind the reader that equation~\NewDM\ for the evolution of the density contrasts in the dark matter assumes all the components are nonrelativistic and all other kinds of matter and energy produce negligible peculiar gravitational accelerations.  However, since~\NewDM\ uses the local physics of relative accelerations it is valid even in the presence of a cosmological constant or of a significant mass density in homogeneously distributed radiation.  In the presence of such terms the mass density $\rho$ in equation~\NewDM\ remains that of the dark matter, and when $\rho$ is significantly smaller than the total source density $\rho_{\rm tot}+p_{\rm tot}$ that is driving the expansion of the universe, the right hand side of \NewDM\ is not a significant source term for the  evolution of the $\delta_q$. This is the situation at redshifts $z\gsim z_{\rm eq}\sim 3000$, where the mass density in radiation and neutrinos dominates the dark matter, and the pressure and self-gravity of the radiation are important.  Since the visible sector is not supposed to be significantly affected by the scalar field, the behavior of our model prior to domination of the expansion rate by the dark matter is very close to $\Lambda$CDM provided the initial scalar charge separation is small.

\subsection{Solutions}
\label{SOLUTIONS}

Let us now solve the linear equations \NewDM\ in the approximation of neglecting all sources of stress energy except the dark matter.  Consider a linear combination of the density contrasts:
 \eqn{DeltaC}{
  \Delta_c = \sum_q c_q \sqrt{f_q} \delta_q \,,
 }
where the $c_q$ are at this point arbitrary constants.  The factor of $\sqrt{f_q}$ is introduced so as to extract from \NewDM\ an eigenvalue problem for a symmetric real matrix, making the completeness properties of the solutions obvious.  One sees from \NewDM\  that the linear combination satisfies
 \eqn{DeltaCevolves}{
  \ddot\Delta_c + {2\dot{a} \over a} \dot\Delta_c - 
   4\pi G \rho 
   \sum_{p,q} c_p \sqrt{f_p} \beta_{pq} f_q \delta_q = 0 \,.
 }
If we choose the $c_q$ so that
 \eqn{ArrangeC}{
  \sum_p c_p \sqrt{f_p} \beta_{pq} \sqrt{f_q} = \xi_c c_q
 }
for some number $\xi_c$, then \DeltaCevolves\ involves $\Delta_c$ only.  This is an eigenvalue problem: if we define
 \eqn{DefineXi}{
  \Xi_{pq} = \sqrt{f_p} \beta_{pq} \sqrt{f_q} \,,
 }
then the $\xi_c$ are the eigenvalues of $\Xi_{pq}$ and the $c_q$ are the corresponding eigenvectors.  Since $\Xi_{pq}$ is real and symmetric, the eigenvectors form a complete set, and we may completely decouple \NewDM\ into equations of the form
 \eqn{SeveralODEs}{
  \ddot\Delta_c + {2\dot{a} \over a} \dot\Delta_c - 
   4\pi G \rho \xi_c \Delta_c = 0 \,.
 }
For $a = (t/t_0)^{2/3}$ (as appropriate for the matter-dominated era), the solutions to \SeveralODEs\ are $\Delta_c \sim t^{2\gamma_{c\pm}/3}$, where
 \eqn{GammaCDef}{
  \gamma_{c\pm} = {-1 \pm \sqrt{1 + 24 \xi_c} \over 4} \,.
 }
When the square root is real, we will set $\gamma_c = \gamma_{c+}$, for the more rapidly growing mode.

Under the condition that the total scalar charge vanishes, so equation \NetChargeVanishes\ applies, one solution to the eigenvalue problem always is the adiabatic mode, $c_q = \sqrt{f_q}$. This is the mass density contrast mode, $\Delta_c=\delta_m$.  Its eigenvalue is $\xi_1=1$, and the rate of growth of the mass density contrast, $\delta_m\propto t^{2/3}$ in the matter-dominated epoch, follows the standard model. The power law indices for the evolution of the isocurvature modes are constrained by the trace of the matrix \DefineXi. Since $\xi_1=1$, the sum of the other eigenvalues is 
 \eqn{TraceofChi}{
\sum_{c>1}\xi_c = {\rm Tr}(\Xi )  - 1 = \sum f_q\beta_{qq} - 1 = 
2\sum_q M_{\rm Pl}^2f_q\left(\sum_sQ^s_q{}^2  - \sum_g Q^g_q{}^2\right) /m_q^2\,.
}
If there are two particle species and the gauge force dominates then the second eigenvalue $\xi_2$ is negative, and the real part of \GammaCDef\ is therefore negative so that the mode amplitude oscillates or decays.  If the scalar force dominates, then there are positive charge separation eigenvalues corresponding to growing mode amplitudes.  
 
These results are illustrated by some examples.
 \begin{itemize}
  \item[1)] Pure CDM: Here there is only one species, and $\beta = f = \xi = \gamma = 1$.  The growth law $\delta_m \sim t^{2/3}$ lies at the heart of the successes of the CDM model.
  \item[2)] Two species with gauge interactions: Here the total force law between particles $1$ and $2$ is
 \eqn{GravityGauge}{
  F = {G m_1 m_2 \over r^2} - {Q_1 Q_2 \over 4\pi r^2} \,,
 }
where $F$ is taken to be positive if the force is attractive, and negative if it is repulsive.  Let us assume that the $q_i$ are all either $+q$ or $-q$, while the $m_i$ are all equal to $m$.  To achieve overall charge neutrality, one must have $f = 1/2$ for both species.  Then
 \eqn{TwoSpeciesGauged}{
  [\beta_{pq}] &= \pmatrix{1 & 1 \cr 1 & 1} - 
   2Q^2 {M_{\rm Pl}^2 \over m^2} \pmatrix{1 & -1 \cr -1 & 1}  \qquad
  [\Xi_{pq}] = {1 \over 2} [\beta_{pq}]  \cr
  \xi_1 &= 1 \qquad \gamma_1 = 1  \cr
  \xi_2 &= -2 Q^2 {M_{\rm Pl}^2 \over m^2} \qquad
   \gamma_{2\pm} = {1 \over 4} \left( -1 \pm
    \sqrt{1 + 24 \xi_2} \right) \,.
 }
If $\xi_2 < 1/24$ the power law indices $\gamma_{2\pm}$ for this mode are imaginary, which corresponds to plasma waves in the visible sector. Since 
$\xi_2 < 0$ the real part of $\gamma_{2\pm}$ is always negative, meaning the amplitude of the charge separation is decreasing. The growing charge-neutral mode, with $\gamma_1 = 1$, thus dominates at late times.
 \item[3)] Two species with scalar interactions: This is the model we will focus on the most.  The total force law between particles $1$ and $2$ is given by \GravityScalar, provided the scalar $\phi$ is canonically normalized.  Let's assume that the mass of one species is $m(\phi)$, the mass of the other is $m(-\phi)$, and $\phi=0$ in the background solution.  One needs $f=1/2$ for both species (scalar charge neutrality) so as not to push $\phi$ away from $0$.  Thus
 \eqn{TwoSpeciesScalar}{
  [\beta_{pq}] &= \pmatrix{1 & 1 \cr 1 & 1} + 
   2 M_{\rm Pl}^2 \left( {d\log m \over d\phi} \right)^2 
    \pmatrix{1 & -1 \cr -1 & 1}  \qquad
  [\Xi_{pq}] = {1 \over 2} [\beta_{pq}]  \cr
  \xi_1 &= 1 \qquad \gamma_1 = 1  \cr
  \xi_2 &= 2 M_{\rm Pl}^2 \left( {d\log m \over d\phi} \right)^2 \qquad
   \gamma_2 = {1 \over 4} \left( -1 +
    \sqrt{1 + 24 \xi_2} \right) \,.
 }
 \item[4)] Four species with gauge and scalar interactions: In an $S^1$ compactification, winding strings carry one conserved gauge charge, while momentum strings carry different one---the gauge group is $U(1) \times U(1)$.  The masses of winding and momentum strings depend oppositely on the scalar (if one is $m(\phi)$, the other is $m(-\phi)$), but they do not depend on the charge.  If $\phi=0$ in the background, corresponding to the self-dual radius, then the winding charge and momentum charge are equal in magnitude (in the sense of giving rise to forces of equal magnitude).  Ordering the species as (winding, anti-winding, momentum, anti-momentum), one arrives at
 \eqn{FourSpecies}{
  [\beta_{pq}] &= \pmatrix{1 & 1 & 1 & 1 \cr 1 & 1 & 1 & 1 \cr
    1 & 1 & 1 & 1 \cr 1 & 1 & 1 & 1} +
    2 M_{\rm \rm Pl}^2 \left( {d\log m \over d\phi} \right)^2
    \pmatrix{1 & 1 & -1 & -1 \cr 1 & 1 & -1 & -1 \cr
     -1 & -1 & 1 & 1 \cr -1 & -1 & 1 & 1} \cr &\qquad{} - 
    2 Q^2 {M_{\rm Pl}^2 \over m^2}
    \pmatrix{1 & -1 & 0 & 0 \cr -1 & 1 & 0 & 0 \cr 0 & 0 & 0 & 0 \cr
     0 & 0 & 0 & 0} - 
    2 Q^2 {M_{\rm Pl}^2 \over m^2}
    \pmatrix{0 & 0 & 0 & 0 \cr 0 & 0 & 0 & 0 \cr 0 & 0 & 1 & -1 \cr
     0 & 0 & -1 & 1} \,,
 }
where we've put gravitational forces in the first term, scalar forces in the second, gauge interactions between winding modes in the third, and gauge interactions between momentum modes in the fourth.  To achieve charge neutrality for the two $U(1)$'s and avoid pushing $\phi$ away from $0$, one needs $f=1/4$ for all species.  Thus
 \eqn{XiEtcFour}{\seqalign{\span\TC}{
  [\Xi_{pq}] = {1 \over 4} [\beta_{pq}]  \cr
  \xi_1 = 1 \qquad 
  \xi_2 = 2 M_{\rm Pl}^2 \left( {d\log m \over d\phi} \right)^2 \qquad
  \xi_3 = \xi_4 = -Q^2 {M_{\rm Pl}^2 \over m^2} \,.
 }}
The first mode is adiabatic (all species move together).  The second retains charge neutrality in both gauge groups, but winding strings cluster in one place while momentum strings cluster in another.  The two remaining modes involve charge separation.  Clearly, these last two modes are subdominant (and may even be oscillatory).  The most interesting physics, then, is the same as in case 3 above.
 \end{itemize}

To summarize, in linear perturbation theory there always is an adiabatic mode in which the mass density fluctuations evolve in the standard way. Gauge interactions tend to suppress the charge separation modes, but the scalar interaction can cause the fractional charge separation to evolve faster than the mass density contrast. The dynamical reason for this is that gauge interactions tend to be screened by the plasma of charged particles, whereas scalars, as our results demonstrate, can be anti-screened, leading to new instabilities.  We note in section~\ref{OBSERVATIONS} that if the charge separation mode amplitudes exceeded that of the mass it could produce observable mass concentrations at positions that are not correlated with the large-scale mass distribution. This effect is not observed, but it is easily avoided by postulating that the initial conditions are close to adiabatic, because the charge separation modes have been growing only since the expansion rate became matter-dominated at $z_{\rm eq}\sim 3000$.  For example, we will show in section~\ref{CHARGES} that in a particularly simple string theory setup, the adiabatic mode grows as $t^{2/3}$ while the scalar charge separation mode grows as $t$.  One can avoid potential conflicts with observations of large scale structure by letting $\delta$ for the charge separation mode be less than a third of $\delta$ for the adiabatic mode.  This corresponds to suppressing the charge separation $\delta$ by a factor of $\sim 100$ at $z \sim 3000$.

We have continued using string-inspired terminology in the examples listed above, but clearly our setup is a simple generalization of the CDM model: the $\beta$ parameters effectively lift the assumption of the universality of free fall in the dark sector, an effect studied in many papers \cite{fp}. 

\subsection{Guidance on Charges and Masses from String Theory}
\label{CHARGES}

Let us now consider the simplest string theory setup and actually compute the $\beta$ parameters.  Reducing the low-energy effective action of string theory from ten dimensions to four on a compact manifold $K$ includes the following manipulation (see for example \cite{ms}):
 \eqn{TenToFour}{
  & {1 \over 2\kappa_{10}^2} 
   \int d^4 x \int_K d^6 y \sqrt{G_{10}} e^{-2\Phi_{10}}
    \left[ R_{10} + 4 (\partial\Phi_{10})^2 \right] \cr 
  &\qquad{} = {1 \over 2\kappa_4^2} 
   \int d^4 x \sqrt{G_4} e^{-\Phi_4} \left[ R_4 + (\partial\Phi_4)^2 + 
    {1 \over 4} \partial_i G_{uv} \partial^i G^{uv} \right]  \cr
  &\qquad{} = {1 \over 16\pi G} \int d^4 x \sqrt{g}
   \left[ R - {1 \over 2} (\partial\Phi_4)^2 + 
    {1 \over 4} \partial_i G_{uv} \partial^i G^{uv} \right]  
    \cr\noalign{\vskip1\jot}
  & e^{-\Phi_4} = {\Vol K \over (2\pi\sqrt{\alpha'})^6} e^{-2\Phi_{10}}
   \qquad g_{ij} = e^{-\Phi_4} G_{ij}  \cr
  & 2 \kappa_{10}^2 = {(2\pi\sqrt{\alpha'})^8 \over 2\pi} g_s^2 \qquad
  2 \kappa_4^2 = 16\pi G = 2\pi\alpha' g_s^2 \,.
 }
The string metric $G_{MN}$ is not rescaled in reducing from ten dimensions to four: the $G_{ij}$ are its non-compact components, and the $G_{uv}$ are its compact components.  The scale of compactification is close to the Planck scale, so it makes sense to let all the fields in \TenToFour\ be independent of the directions in $K$.  The term $\partial_i G_{uv} \partial^i G^{uv}$ in the action summarizes massless scalar moduli.  There are other massless moduli at tree level, coming for instance from the anti-symmetric tensor field $B_{MN}$.

Let us assume than in the unperturbed solution, $K$ is a square $T^6$ whose constituent circles all have circumference $2\pi\sqrt{\alpha'}$.\footnote{$T^6$ is actually not a suitable compactification manifold because it leads to ${\cal N}=8$ supersymmetry in four dimensions, which does not admit chiral fermions and cannot be broken spontaneously.  Working through this example, however, gives a good sense of why the $\beta$ parameters are of order unity.}  Let the six coordinates $y^u$ run from $0$ to $2\pi\sqrt{\alpha'}$, so that $G_{uv} = \delta_{uv}$.  Now consider perturbing a single diagonal entry: $G_{99} = e^{2\phi/M_{\rm Pl}}$, where as usual $M_{\rm Pl}^{-2} = 8\pi G$ in four dimensions.  The four-dimensional action in Einstein frame includes the terms
 \eqn{FourDTerms}{
  S = \int d^4 x \sqrt{g} \left[ {1 \over 16\pi G} R - 
   {1 \over 2} (\partial\phi)^2 \right] \,.
 }
The circle in the $9$ direction has radius $R = \sqrt{\alpha'} e^{\phi/M_{\rm Pl}}$.  The lowest mass Kaluza-Klein state with momentum in the $9$ direction has mass $m_1 = 1/R$, and the lowest mass string wrapping the $9$ direction has mass $m_2 = R/\alpha'$;\fixit{Need to check these expressions} thus
 \eqn{StringMasses}{\seqalign{\span\TC}{
  m_1 \sqrt{\alpha'} = e^{\phi/M_{\rm Pl}} \qquad
   m_2 \sqrt{\alpha'} = e^{-\phi/M_{\rm Pl}} \cr
  [\beta_{pq}] = \pmatrix{3 & -1 \cr -1 & 3} \,.
 }}
Scalar charge neutrality gives $f_1 = f_2 = 1/2$, so $\xi_1 = 1$ for the adiabatic mode and $\xi_2 = 2$ for the charge separation mode.\footnote{If there is an extra species of non-relativistic matter with mass fraction $f_3$ and no coupling to the scalar field (for example, baryonic matter), then $\xi_2 \to \xi_2 (1-f_3)$ as compared to the situation where this extra species is absent.}  The numbers for the $\beta_{pq}$, the $f_q$, and the $\xi_q$ are all independent of the background value of $\phi$: this is a special feature of the exponential dependence of the masses on $\phi$ in \StringMasses.

A potential difficulty is that in four-dimensional Einstein frame, the effective value of $\alpha'$ depends on the four-dimensional dilaton $\Phi_4$.  The presence of wrapping and winding strings will tend to drive $\Phi_4$ toward more negative values (weak coupling).  For the purposes of our simple-minded discussion, we might simply assume that $\Phi_4$ is stabilized by some unspecified mechanism.  Or we could suppose that there are D-branes wrapping extra dimensions to stabilize $\Phi_4$.

The values of the $\beta_{pq}$, and hence of $\xi_2$, depend on the normalization of the scalar kinetic term.  Such normalizations are difficult to compute reliably in string theory because they depend on the Kahler potential, a quantity which is not protected from renormalization even in the presence of ${\cal N}=1$ supersymmetry (it is protected by ${\cal N}=2$ supersymmetry).  One may also object that the model we have used for \TenToFour\ is excessively naive.  Nevertheless, string theory provides a clear motivation for the scalar forces to be roughly as strong as gravitational forces: these two forces have the same origin in the higher-dimensional description.

More generally, what is needed to have $\xi_2$ on the order of unity is for the masses of dark sector particles to have the form 
 \eqn{mqForm}{
  m_q = m_{q0} f_q(\phi/M_{\rm Pl}) \,,
 }
where $f_q$ is a function whose values and derivatives are order unity and otherwise unconstrained.  The value of the constant $m_{q0}$ is not important.  Thus we are not tied to dark matter particles with masses near the Planck scale: if there is a mechanism which explains the smallness of $m_{q0}$, then the form \mqForm\ is natural for $\phi$ regarded as a modulus of some family of vacua.\footnote{To an extent, \StringMasses\ is an example of \mqForm: if the string coupling is very weak, then $1/\sqrt{\alpha'} \ll M_{\rm Pl}$.}

\section{Constraints from astronomy}
\label{OBSERVATIONS}

A viable alternative to the standard $\Lambda$CDM cosmology must pass a considerable and quite demanding suite of cosmological tests \cite{Bennettetal}. To satisfy these tests we follow the standard model in assuming the presence of a cosmological constant, $\Lambda$, or a term in the stress-energy tensor the acts like it. The values of the dark matter particle masses and scalar charges in our model are determined by the dynamics of a scalar field, but the field value has been locked---apart from the small spatial gradients driven by the inhomogeneous dark matter distribution---at redshifts well beyond those of observational interest. The mean expansion history thus is observationally indistinguishable from the standard cosmology. To avoid unwanted (because unobserved) effects on the interaction of  the distributions of mass and radiation at decoupling, we must assume the scalar charge separation is small at decoupling. We satisfy this by assuming close to adiabatic initial conditions. Under all these assumptions our model passes the cosmological tests based on observations that probe the mean expansion history of the universe and the behavior of density fluctuations on scales comparable to the present Hubble length. 

The observation that galaxies are useful tracers of mass (see \cite{Tegmark:2003uf} and references therein) requires us to stipulate that the initial conditions are close enough to adiabatic on the scale of galaxies that the amplitude of the charge separation mode was small compared to the perturbation to the total mass distribution as the galaxies were forming. This is because an initially isocurvature mode, with charge density contrasts $\delta_-=-\delta_+$ and no perturbation to the total mass density, becomes a mass fluctuation as it grows nonlinear,  $\delta_-$ is driven to zero,  and $\delta_+$ grows to values much greater than unity. The result is a mass concentration---usually termed a massive halo---compensated by a surrounding hole. 

 \begin{figure}
  \vskip-0.3in
  \centerline{\includegraphics[width=3.3in]{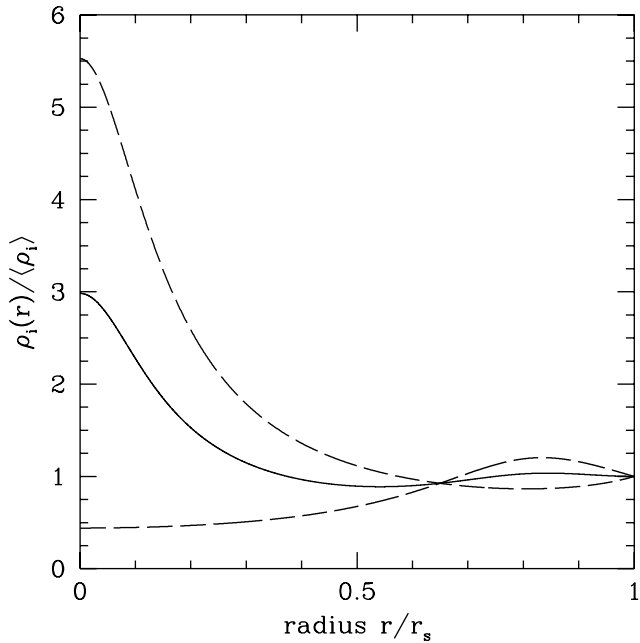}}
  \caption{An illustration of the development of a massive halo out of an initially isocurvature charge perturbation, that is, an initially homogeneous mass distribution with a small scalar charge separation. The system is spherically symmetric. The dashed curves are the densities relative to the mean for the two dark matter species, and the solid curve is the total mass density.  All curves are plotted at the moment when the dense component has just stopped expanding in the center.}\label{fig2}
 \end{figure}

An example of this effect might be useful. The simple illustration in figure~\ref{fig2} assumes spherical symmetry and two types of dark matter with the same mean mass density and $\beta_{++} = \beta_{--} =2$,  $\beta_{+-}=0$. This means that the force of attraction of like charges is twice that of standard physics and the scalar repulsion of unlike charges just cancels the attraction of gravity. The initial density contrasts in the growing modes of the two dark matter components are
\eqn{delta_is}
{N_\pm (<\!r)/\bar nV(<\!r) = 1 \pm\delta_i \cos^2(\pi r/2R),
}
where $N_\pm (<\!r)$ and $V(<\!r)$ are the particle number and volume within radius $r$, and $\bar{n}$ is the mean number density. This produces density contrasts $\delta_\pm = \pm\delta_i$ at the center of the system, while at  the outer radius $R$ the contained mass is the same as in the background cosmological model. The expansion at the outer radius thus follows the background model, which is Einstein-de Sitter in this example. The initial value of the perturbation $\delta_i$ at the center is chosen so that after the outer radius of the system has expanded by a factor of 30 the central part of the denser component has just stopped expanding and is about to start collapsing. The figure shows the density runs of the two dark matter components at this time, when the central densities are $\rho_+/\bar\rho_+ = (3\pi /4)^2$ and $\rho_-/\bar\rho_- =0.442$. The solid curve shows the total mass density, with a mass concentration at the center and a compensating mass deficit near the edge of the system. The lesson is that massive halos can grow out of a primeval homogeneous mass distribution with scalar charge density fluctuations through the nonlinear development of the charge separation. 

If on the scale of galaxies the primeval charge separation fluctuations extrapolated to the present epoch in linear perturbation theory were comparable to the primeval mass fluctuations it would produce massive halos at positions that are not closely correlated with the large-scale mass distribution. Perhaps the tightest constraint on this effect is the observation that no known type of object tends to populate the voids defined by the concentrations of normal galaxies with luminosities comparable to the Milky Way \cite{peeb02}. Also striking is the consistency between the measured anisotropy of the 3~K thermal background radiation and what is predicted by initially adiabatic mass density fluctuations that are traced now by the galaxies (\cite{Bennettetal}, \cite{Tegmark:2003uf} and references therein). Thus we must  postulate that  the initial conditions are close to adiabatic on scales greater than and comparable to galaxies.

Under pure adiabatic initial conditions, gravity drives the formation of massive halos that are well mixed and fair samples of all the pressureless dark matter components. A well-mixed and neutral massive halo can be unstable against separation of the scalar charges, however. A particularly simple case, for two charge components, is $\beta_{++}=2$, $\beta_{+-}=0$. This means the gravitational and scalar forces between oppositely charged components just cancel, while the force of attraction of like charges is twice that of gravity. A positive and negative massive halo thus would interact with each other only by the gravity of the visible sector mass of  gas and stars that pulls on both. Massive halos of the two types would tend to form at the same place, out of the same primeval mass distribution, but, depending on the relative sizes of the isocurvature and adiabatic initial conditions, and on the mass in the visible sector, initially overlapping halos may drift apart.  

At larger scalar charge-to-mass ratio, charge separation is driven by a dynamical instability, as one sees by the following consideration. Imagine a bound spherical halo of two species of dark matter particles with masses $m_+$ and $m_-=\mu m_+$, with  $\mu\leq 1$. There are equal numbers of particles with scalar charges $Q_-=-Q_+$, so the halo has no net charge. We will suppose that the two charge distributions behave as two continuous fluids, and that the two fluids have close to the same initial distributions so the total scalar charge density is close to zero everywhere. Let the characteristic radius of the fluid distributions be $r$. Under slow adjustments of $R$ the energy of the system varies with the radius as
\eqn{haloa}{
E = a(1+\mu )/r^2 - b(1+\mu )^2/r,
}
where $a$ and $b$ are constants. The first term on the right-hand side is the kinetic energy. As indicated, the rms velocity scales inversely as $r$ under a uniform adiabatic change of radius. The second term is the gravitational potential energy, which is proportional to the square of the mass and scales inversely as the radius. The minimum of the energy is at 
\eqn{halob}{
r = 2a/b(1+\mu ),\qquad E = - b^2(1+\mu )^3/4a .
}
Let us compare this to the energy of the system when one fluid has expanded to a much larger radius; the relevant case is where the less massive component has moved to much lower density. Now the  energy of the system is dominated by that of the more massive component, so 
\eqn{haloc}{
E_1 = a/r_1^2 - \beta_{++}b/r_1,
}
when the characteristic radius of the more massive component is $r_1$. The mass that now dominates the energy is down by the factor $1+\mu$ from the first case, and the gravitational binding energy is increased by the scalar interaction, which is represented  by the factor $\beta_{++}$ defined in equation~\BetaDefine . The energy $E_1$ energy is minimum at
\eqn{halod}{
r_1 = 2a/\beta_{++}b, \qquad E_1 = - b^2\beta_{++}^2/4a .
}
The charge separated state is energetically favored if
\eqn{haloe}{
\beta_{++} > (1+\mu )^{3/2}.
}

We do not know an easy way to see whether the well-mixed state in this example has a local energy minimum that would slow the growth of charge separation. Our highly simplified analysis might be expected to underestimate the instability by ignoring the binding energy of the more dispersed dark matter; a similar though even more crude argument suggests that a near spherical halo with radially separated scalar charges is energetically disfavored against the formation of two separated halos of opposite signs, one less tightly bound than the other.  Our conclusion is that if $\beta_{++}$ is significantly greater than unity, then well-relaxed massive halos have a single sign of the dark matter. 

It should be noted that the physics does not set a length scale for this charge separation instability: that depends on size of the strongly nonlinear mass concentrations. Thus charge separation might be complete within the effective radii (which contain half the starlight) of galaxies, charge separation might be in progress on the  scale of clusters of galaxies, and it likely would not yet be happening on larger scales. In particular, the large-scale galaxy flows would be driven by gravity alone, as in the standard model. Also, under hierarchical structure formation we would not expect the formation of well-mixed halos of dark matter with $\beta_{++}\gsim 1$ on the scale of present-day galaxies. Rather, charge separation would commence in the first generation of bound dark matter halos, and subsequent merging and accretion would favor the growth of separately charged massive halos, likely with envelopes of more diffuse dark matter with the opposite scalar charge. 

We conclude this section with a few comments on what might learned from numerical N-body explorations of how structure would form in this model, and on the possible guidance one might find in the phenomenology. We consider the case  $2\lsim \beta_{++}\lsim 4$, where the scalar force between unshielded charged halos is important but not very much larger than gravity. 

The smoothly distributed dark matter in a cluster of galaxies might be dominated by one sign of the scalar charge at the center and the other sign on the outskirts. That would not discourage gravitationally-driven merging of clusters, but it might encourage persistent separation into two mass concentrations. It would drive the separation of galaxies in a cluster that have their own differently charged massive halos, and it would tend to push the stars in a galaxy away from the center of the galaxy massive halo, a possibly observable effect. 

A charge-separated massive halo of a galaxy would be about twice as tightly bound as in standard physics, for given dark matter mass and radius. The effect would be observable only indirectly, in differences in the motions of satellites that have their own charged dark matter halos and the motions of gas and stars that are not bound to satellite halos.  

The nearest galaxy to the Milky Way, the Andromeda Nebula, is falling toward us, likely for the first time. It is likely therefore that both of these dominant members of the Local Group formed in relative isolation, so each has close to zero net scalar charge, with a massive halo of one sign and an envelope of the other. This would mean that the galaxies are approaching under the usual gravitational attraction, and the same would be true of the irregular galaxies on the outskirts of the Local Group. The galaxy NGC 6822 is about 500~kpc from the Milky Way, about half the distance of the Andromeda Nebula, and possibly in a region where its orbit has been influenced by the scalar force of dark matter that has been pushed away from the cores of the two galaxies. We will be watching with interest closer analyses of its orbit, as may be possible with new generations of precision astrometry. 

Pairs of galaxies with oppositely charged massive halos would not merge.  Pairs of galaxies with separations comparable to their effective radii are not uncommon: there is a significant number in the rich cluster MS1054, at $z=0.8$ \cite{MS1054paper}; Hu and Cowie \cite{CowieHu} find a curious number of close pairs with close redshifts in a field at $z\sim 0.4$; Struble and Rood \cite{StrubleRood} suggest some Abell clusters at low redshift have significant numbers of close pairs; binary giant galaxies at the centers of clusters of galaxies are not uncommon \cite{MorganLesh,TodLauer}; and, finally, there are enough close pairs of galaxies in the field at low redshift to make the galaxy two-point correlation function a good approximation to a power law down to separations $r \sim 10$~kpc, comparable to typical effective radii, meaning the visible parts of the galaxies are close to touching. In the standard model a close pair of galaxies soon merges because of the gravitational dynamical drag of the overlapping dark halos. The multiple nuclei of brightest cluster galaxies may be remnants of recent mergers of clusters, but the idea that the large number of close galaxy pairs in MS1054 formed this way is curious because the relative velocities of the galaxies are much larger than the velocities within galaxies. It is conceivable also that the orbits of bound systems of galaxies in the field decay in just such a way as to preserve the power law two-point correlation function, but that seems to be a curious conspiracy. The scalar charge model offers a new line of speculation: perhaps close galaxy pairs with the same sign of dark matter have long since merged, while pairs with opposite signs of the scalar charge remain, loosely bound by the gravitational attraction of their baryons, and long-lived because the dominant drag is from the baryonic matter. The concept is curious, but so is the situation. 

We continue this speculative discussion in \cite{gpTwo}, where the scalar interaction in the dark sector has some different features.

\section{Naturalness, coupling to the visible sector, and the weak equivalence principle}
\label{NATURAL}

In order to significantly affect galaxy formation, the range of the scalar force needs to be at least on the order of a megaparsec.  The corresponding mass is unnaturally small: $m^2\phi^2$ is a relevant deformation of the lagrangian, and barring some symmetry one would generically expect $m$ to be on the order of the Planck scale.  It won't do to have $\phi$ be a Goldstone boson, because if it were, the non-derivative tree-level couplings we need to the dark matter particles would be forbidden.  Supersymmetry is another possibility for preventing $\phi$ from acquiring a mass through loop effects.  But supersymmetry, if present at all in Nature, is broken at the ${\rm TeV}$ scale in the visible sector, and based on the conventional understanding of supersymmetry breaking, it is hard to see how it would be preserved to the scale of inverse megaparsecs in the dark sector.  To be more precise, assume that the fundamental scale of supersymmetry breaking is $10 \, {\rm TeV}$, with gauge mediation to the visible sector and only gravity mediation to the weak sector.  Then based on the conventional understanding of supersymmetry breaking, the mass of $\phi$ from a soft supersymmetry breaking term is on the same order as the mass of the gravitino---about $0.1 \, {\rm eV}$.  This is still too big by many orders of magnitude.  We do not have a solution to this apparent fine-tuning problem, which applies equally to this paper and to \cite{gpTwo} (not to mention the vast additional literature on light scalar fields in various contexts).

We are however disinclined to rule out a model on grounds that naturalness is violated only after supersymmetry breaking.  After all, the smallness of the cosmological constant is precisely such a violation.

In the following subsections we will argue that before supersymmetry breaking, a light scalar which does not roll need not be in severe conflict with field-theoretic notions of naturalness.  At first blush this may appear to amount only to the familiar fact that supersymmetric theories can have flat directions.  But supergravity corrections to Yukawa couplings lead inevitably to couplings of $\phi$ to the visible sector, and our main purpose is to show that these couplings need not spoil standard tests of the equivalence principle, provided the superpotential and Kahler potential have certain properties.  In \cite{gpTwo} we will discuss the extent to which these properties can be motivated from string theory.  A brief preview of the string theory motivation will be given in section~\ref{GRABBAG}.

\subsection{Coupling to the visible sector}
\label{VISIBLE}

The crucial point regarding coupling to the visible sector is that couplings should start at dimension six, with a form suggested by supergravity, suppressed by a factor of $1/M_{\rm Pl}^2$.

The dimension six operators in question include a factor of $\phi^2$, so $\phi$ must sit close to $0$ in the background solution---otherwise there is a dimension five component to the visible sector couplings which would spoil the argument.  This may seem to be a fine-tuning of the state of the universe,\footnote{See however \cite{dp} for an elegant self-tuning mechanism.} particularly with dark matter species present whose mass is not minimized at $\phi=0$.  There is less difficulty when the scalar force is screened by the effects of an additional dark matter species which is massless at $\phi=0$, as proposed in \cite{fp}.  Further discussion of this point is deferred to \cite{gpTwo}.

If $\phi$ comes from the size of a compact dimension, such as the $S^1$ in the $x^9$ direction considered in section~\ref{CHARGES}, then it is not clear that couplings to the visible sector starting at $\phi^2$ can be arranged.  We will comment further upon this in section~\ref{GRABBAG}, and we will explore an alternative geometric origin of $\phi$ in \cite{gpTwo} where such coupling is better motivated.  For now, we will simply assume the minimal couplings of the scalar to the visible sector allowed by low-energy supergravity. 

In the framework of $d=4$ ${\cal N}=1$ supergravity, these minimal couplings are non-renormalizable couplings in Yukawa interactions and $F$-term contributions to the scalar potential.  It is possible for there to be no such couplings in gauge interactions and $D$-term contributions to the scalar potential.  The underlying reason is that the superpotential is a section of a line bundle over the Kahler manifold parametrized by the chiral superfields, so there are factors of $e^{K/2 M_{\rm Pl}^2}$ in front of each Yukawa couplings, and similarly an overall factor of $e^{K/M_{\rm Pl}^2}$ in the scalar potential; on the other hand, complexified gauge couplings are genuinely holomorphic functions of the chiral superfields, so gauge interactions descend from the superspace action in a way that does not involve the Kahler potential.  We assume that the dependence of $K$ on the chiral superfield that includes the scalar $\phi$ is $K = \Phi^\dagger \Phi + \ldots$: this gives rise to a canonically normalized kinetic term for $\phi$.  It is also necessary to suppress dimension four terms in the superpotential which are linear in $\phi$ and couple to the visible sector: these would give rise to dimension five couplings in the lagrangian linear in $\phi$.

In the framework described in the previous paragraph, nucleon masses will depend on the modulus $\phi$ through the mass dependence of the constituent quarks.  For instance, for the up quark and the proton,
 \eqn{MassDeviations}{
  m_u &= \bar{m}_u + \epsilon_u {\phi^2 \over M_{\rm Pl}^2}  \cr
  m_p &= \bar{m}_p + \epsilon_p {\phi^2 \over M_{\rm Pl}^2} \,,
 }
where naively, $\epsilon_p = 2\epsilon_u + \epsilon_d$.  All these $\epsilon$ coefficients are expected to be on the order of $m_u$ and $m_d$, that is, $5$ to $10\,{\rm MeV}$.  The analogous coefficient $\epsilon_n$ for the neutron will be different from $\epsilon_p$ because the quark content is different.  This dependence of nucleon masses on a scalar is similar to the situation for variations of the fine-structure constant, except that the mass shifts depend quadratically on $\phi$.  In the next subsection we will replay some standard arguments to set limits on $\langle \phi \rangle$.  These limits will also tell us the accuracy to which we need the Kahler potential and superpotential to satisfy the properties described above which lead to \MassDeviations.  The couplings resulting in \MassDeviations\ as well as some of the estimates to follow are reminiscent of the discussion of a ``Least coupling principle'' in \cite{dp}.

\subsection{The weak equivalence principle}
\label{EQUIVALENCE}

Assuming the dependence \MassDeviations, if $\phi=0$ in the background, an exchange of two scalars in a loop diagram produces an attraction between two nuclei (or more generally between visible objects) which is completely negligible in comparison to gravity:
 \eqn{ForceComparison}{
  F_{\rm loop} = {3\pi^5 \over 8} 
   {\epsilon^2 \over M_{\rm Pl}^4 r^4} \qquad
  F_g = {m_p^2 \over 8\pi M_{\rm Pl}^2 r^2} \qquad
  {F_{\rm loop} \over F_g} = 3 \pi^6 {\epsilon^2 \over
   m_p^2 M_{\rm Pl}^2 r^2} \,,
 }
and the last expression is small for $r \gsim 1/3M_{\rm Pl}$.  If, however, $\langle\phi\rangle \neq 0$, then there is a tree-level coupling of the nucleon to the scalar: effectively,
 \eqn{WeakLinear}{
  m_p = \bar{m}_p + 2\epsilon_p 
   {\langle\phi\rangle \delta\phi \over M_{\rm Pl}^2} \,,
 }
and this gives rise to a force which can again be compared with gravity:
 \eqn{LinearCompare}{
  F_s = {\epsilon_p^2 \langle\phi\rangle^2/M_{\rm Pl}^4 \over \pi r^2} \qquad
  \eta \equiv {F_s \over F_g} = 8 {\epsilon_p^2 \over m_p^2} 
   {\langle\phi\rangle^2 \over M_{\rm Pl}^2} = 8 {\epsilon_p^2 \over m_p^2}
   {\langle\phi\rangle^2 \over M_{\rm Pl}^2} \,.
 }
Recall that the $\epsilon$ coefficients and hence the scalar forces would differ by factors of order unity for protons versus neutrons.  Thus the scalar force would be isotope dependent.  Tests of the universality of free fall put strong constraints on such forces, which roughly amount to $\eta \lsim 10^{-12}$.  This translates to $\langle\phi\rangle \lsim 5 \times 10^{-5} M_{\rm Pl}$.

Spatial variation of $\phi$ could also modify gravitational effects at a measurable level.  Assume there is a spatial gradient of $\phi$ in the Solar System.  The Sun and the Earth, having a different fraction of protons to neutrons, will feel the gradient differently, so the scalar will make different contributions to their total accelerations.  The difference $(\Delta a)_s$ between their total accelerations should be very much less than the acceleration of the earth due to the sun's gravity, $(\Delta a)_g$:
 \eqn{Acceleration}{
  (\Delta a)_s \sim {\epsilon \over m_p} \left| \nabla 
   {\phi^2 \over M_{\rm Pl}^2} \right| \qquad
  (\Delta a)_g = {G M_{\rm sun} \over R^2_{\rm sun}} = 
   {v_{\rm earth}^2 \over R_{\rm sun}} \approx 
   {1 \over 2000 \, {\rm yrs}} \,.
 }
Taking $(\Delta a)_s / (\Delta a)_g \lsim 10^{-13}$ as a rough guide to experimental bounds \cite{AndersonWilliams}, we find that
 \eqn{PhiVariation}{
  \left| \nabla {\phi^2 \over M_{\rm Pl}^2} \right| \lsim 
   H_0/10^4 \,,
 }
where $H_0$ is the present day Hubble constant: $H_0^{-1} \sim 4000 \, {\rm Mpc}$.  If we assume that a characteristic scale of variation for the scalar is the size of the galaxy, around $10 \, {\rm kpc}$, we arrive at a variation of $\phi$ across the galaxy on the order
 \eqn{PVgalactic}{
  (\Delta\phi)_{\rm galaxy} \sim 
   \sqrt{R_{\rm galaxy} |\nabla \phi^2|} \lsim 
    2 \times 10^{-5} M_{\rm Pl} \,.
 }

In fact, in our model, scalar charge separation should have developed within the halo of the Milky Way, and as a consequence, $\langle\phi\rangle$ in the galaxy and $\nabla\phi$ are non-zero.  To estimate how big they should be, note that with all coefficients $\beta_{pq}$ of order unity, the acceleration of a test dark matter particle at the Earth's location due to the halo should be roughly equal to the Earth's average acceleration toward the galactic core, which is $v^2/R_{\rm galaxy}$ with $v \approx 200 \, {\rm km/s}$.  But a dark matter particle's acceleration can be estimated as
 \eqn{DarkAccelerate}{
  a_s = {1 \over m} {dm \over d\phi} |\nabla\phi| =
   {|\nabla\phi| \over M_{\rm Pl}} \sim {1 \over R_{\rm galaxy}}
   {\langle\phi\rangle \over M_{\rm Pl}} \,.
 }
This leads to $\langle\phi\rangle / M_{\rm Pl} \sim v^2/c^2 \approx 5 \times 10^{-7}$, comfortably within the limits arrived at just after \LinearCompare\ and in \PVgalactic.

If there is some linear dependence of quark masses on $\phi$---say a term $c_1 \epsilon_u \phi/M_{\rm Pl}$ added to the right side of \MassDeviations---then one can rework the above estimates to bound $c_1$.  Its effects are similar to the effects of non-zero $\langle\phi\rangle / M_{\rm Pl}$, and the bounds come out to $|c_1| \lsim 2 \times 10^{-5}$.  So for example, a term in the superpotential of the form ${1 \over M_{\rm Pl}} \Phi Y_u H_u Q_L u_R$ would have to be suppressed by an additional factor of $c_1$ to avoid causing problems.

\subsection{Other particle physics considerations}
\label{GRABBAG}

There is a long list of additional issues of potential interest or concern for the model we have presented.  Here we will consider only two.

First, there is a disconnect between the original string theory motivation, where the scalar $\phi$ is the modulus controlling the size of an extra dimension, and the optimistic assumptions about the suppression of couplings of $\phi$ to the visible sector.  Conventional string compactifications give rise to some dependence of the grand-unified gauge couplings on the moduli.

What we have in mind as a more plausible picture of $\phi$ in string theory is a modulus which pertains to dynamics which is localized on the compactification manifold.  For example, if a compactification includes a D3-brane on top of D7-branes, then before supersymmetry breaking there is a flat direction corresponding to moving the D3-brane off the D7-branes.  If the visible sector is assumed to arise from dynamics localized elsewhere on the compactification manifold, then couplings of the dark sector scalars to the visible sector will indeed be suppressed --- at least at tree level.  In \cite{gpTwo} we  describe more fully a class of supersymmetric field theories which lead to interesting dark sector dynamics that includes scalar-mediated forces.  We will also revisit in \cite{gpTwo} the troublesome issue of supersymmetry breaking.

An additional subtlety is that dependence of heavy quark masses on the scalar, as in \MassDeviations, indirectly affects the low-energy gauge couplings, because it changes the location of the threshold between one behavior of the running couplings and another.  We have not given careful consideration to the bounds one might place on couplings to the visible sector through this indirect effect.

Second, cold dark matter coming from massive states of string/M theory cannot be produced as a thermal relic.  If inflation is to be part of the model, then alternative production mechanisms must be considered  (see for example \cite{ckr} and references therein).  Such mechanisms allow dark matter particles to be much more massive than the weak scale.  Such particles would not be detected in experiments designed to detect LSP-like WIMPs.  Indeed, it is difficult to see how one could naturally couple $\phi$ at tree level to ordinary WIMPs without at the same time coupling it unacceptably to the visible sector.

Direct detection of LSP-like WIMPs would of course be encouraging to the view that the dark matter is comprised of such particles.  But until there is a way to experimentally estimate the density of such particles, the possibility remains open that they are at most a subdominant component of dark matter, and that more massive and/or more exotic objects are the dominant component.  As long as this possibility exists, it seems worthwhile to explore possibilities motivated from string theory or in other ways.

\section*{Acknowledgments}

We are grateful to Neal Dalal, Glennys Farrar, David Lyth, Indrajit Mitra, and Scott Watson for useful discussions, and to Robert Brandenberger for commenting on a draft of the manuscript.  The work of SSG~was supported in part by the Department of Energy under Grant No.\ DE-FG02-91ER40671, and by the Sloan Foundation.  SSG also thanks the KITP at UCSB for hospitality; his work there on this paper was supported in part by the National Science Foundation under Grant No.~PHY99-07949.

\newpage

\section*{Appendix}

The purpose of this appendix is to start with the action
 \eqn{ActionAgain}{
  S = \int d^4 x \sqrt{g} \left[ {R \over 16\pi G} - 
   {1 \over 2} (\partial\phi)^2 - V(\phi) \right] - 
   \sum_\alpha \int_{\gamma_\alpha} ds \, m_\alpha(\phi)
 }
and derive the appropriate special case of \NewDM\ as a late time limit, assuming $V(\phi)=0$.  Working through this explicitly will reveal an additional condition \eno{meffStable} on the functions $m_q(\phi)$ that gives stability at early times, and it will hopefully serve to orient readers unfamiliar with density perturbation calculations.  We will following the conventions of \cite{PeeblesSecondBook} except for metric signature, which we take to be mostly plus.

To treat the particles in the hydrodynamic approximation, we introduce the average velocities, the number densities and currents, and the stress tensors associated with each species:
 \eqn{Currents}{
  N_q^i = n_q U_q^i \qquad T_q^{ij} = n_q m_q(\phi) U_q^i U_q^j \,.
 }
The velocities are subject to the normalization condition $U_q^2 = -1$.  The stress tensor for the scalar is
 \eqn{ScalarT}{
  T_s^{ij} = \partial^i \phi \partial^j \phi - g^{ij}
   \left( {1 \over 2} (\partial\phi)^2 + V \right) \,.
 }
The following equations dictate the dynamics of the fluids coupled to the scalar field and gravity:
 \eqn{eoms}{
  &\nabla_i N_q^i = 0 \qquad \nabla_i T^i_{qj} = -n_q \partial_j m_q  \cr
  &\square\phi = {dV \over d\phi} + \sum_q n_q {dm_q \over d\phi} \qquad
   R_{ij} = 8\pi G \left( T_{ij} - {1 \over 2} g_{ij} T^k_k \right) \,.
 }
Here $T^{ij}$ is of course the sum of the $T_q^{ij}$ and $T_s^{ij}$.\footnote{It takes a bit of thought to convince oneself of the second of these equations ($\nabla_i T^i_{qj} = -n_q \partial_j m_q$).  A quick intuitive check is that for a static homogenous fluid in flat space (with gravity turned off), the $j=0$ component of this equation states correctly that $-\dot\rho_q = -n_q \partial_0 m_q$.  The equation we have written is the unique one which gives the correct result in this case, is generally covariant, and results in a total stress tensor $T^{ij}$ which is conserved when the scalar equation of motion is satisfied.}

Consider linear perturbations around a $k=0$ FRW background:
 \eqn{PerturbDeltaPhi}{
  \phi = \bar\phi + \delta\phi \qquad n_q = \bar{n}_q (1+\delta_q) \,.
 }
We will assume that the background solution has $\bar\phi = 0$ for all time, so $\phi = \delta\phi$.  This can only happen if $dV/d\phi = 0$ and $\sum_q n_q dm_q/d\phi = 0$ at $\phi=0$.  The background is matter-dominated FRW provided we assume that also $V=0$ at $\phi=0$.

The perturbations \PerturbDeltaPhi\ couple at linear order to perturbations of the metric, but there is a simple way of decoupling them: we may consider separately the cases where the total matter density perturbation, $\delta_m = \sum_q f_q \delta_q$, vanishes, and the case where $\delta_q = \delta_m$ for all $q$.  In the latter case (the adiabatic mode), it is straightforward to check that the scalar perturbation may be set to $0$, and then \eoms\ reduces to the standard equations of the CDM model.  Derivations of \KeyCDM, which is an exact consequence of \eoms\ for the adiabatic mode, can be found in standard references, and we will not repeat the derivation.

Assuming then that $\delta_m = \sum_q f_q \delta_q = 0$, we first observe that the perturbation in $T^{ij}$ vanishes at linear order.  This may be intuitively obvious, but let us review the argument.  The scalar contribution to $T^{ij}$ is second order because $T_s^{ij}$ is quadratic in $\phi$.  So
 \eqn{deltaT}{
  \delta T^{ij} = \sum_q \delta T_q^{ij} = 
   \sum_q \left[ \delta (n_q m_q) U_q^i U_q^j + 
    n_q m_q (\delta_q U_q^i U_q^j + U_q^i \delta U_q^j) \right] \,.
 }
In the background solution, $U_q^\alpha = 0$ for all spatial components $\alpha$ and all species $q$.  Perturbations $\delta U_q^i$ are allowed to be non-zero only for spatial $i$, so as to preserve the norm of $U_q^i$.  Therefore, the second and third terms in square brackets can be non-zero only when one of $i$ and $j$ is $0$ and the other is spatial.  But we may employ gauge freedom to set $\delta T^{i0} = 0$: this is synchronous gauge.  The first term in square brackets can be non-zero only for $i=j=0$, and then it is
 \eqn{OneComponent}{
  \delta T^{00} = \sum_q \delta (n_q m_q) = 
   \sum_q n_q m_q \delta_q + 
   \left( \sum_q n_q {dm_q \over d\phi} \right)
    \delta\phi \,.
 }
In the last expression, the sum in parentheses vanishes by assumption, and the other sum is proportional to $\sum_q \Omega_q \delta_q$, which also vanishes.  Thus we can consistently set $\delta T^{ij} = 0$, as claimed.  It follows that we may also set $\delta g_{ij} = 0$: this is an isocurvature mode.

We now perturb \eoms\ with the metric held fixed.  The conservation equation for $N_q^i$ results in
 \eqn{NumberCL}{
  \dot\delta_q + \partial_\alpha U_q^\alpha = 0 \,,
 }
and $\partial_\alpha$ acting on the equation for $\partial_i T^i_{q\alpha}$ results in
 \eqn{StressCL}{
  U_q^i \partial_i (m_q a^2 \partial_\alpha U^\alpha_q) &= 
   -{dm_q \over d\phi} \partial_\alpha \partial_\alpha \phi  \cr
  \ddot\delta_q + {2\dot{a} \over a} \dot\delta_q &=
   {d\log m_q \over d\phi} \partial_\alpha \partial_\alpha \phi
 }
where to get the second line we have used \NumberCL.  For the scalar equation, $\phi$ itself is a first order quantity since it is zero in the background, so we have
 \eqn{ScalarPerturb}{
  -\ddot\phi - 3 {\dot{a} \over a} \dot\phi + 
   {1 \over a^2} \partial_\alpha \partial_\alpha \phi  = 
   {d^2 V \over d\phi^2} \phi + 
   \rho \sum_q {f_q \over m_q} {d^2 m_q \over d\phi^2} \phi + 
   \rho \sum_q f_q {d\log m_q \over d\phi} \delta_q \,,
 }
where we have used $n_q = \rho f_q/m_q$ in the background solution.  At early times, when $\partial_\alpha \partial_\alpha \phi$ is negligible on the left hand side of \ScalarPerturb, the perturbations are stable provided
 \eqn{meffStable}{
  m_{\rm eff}^2 \equiv {d^2 V \over d\phi^2} + 
   \rho \sum_q {f_q \over m_q} {d^2 m_q \over d\phi^2} > 0 \,.
 }
The numerical study \cite{bw} applies to precisely such modes whose wavelengths are well outside the horizon.

At late times, when $\partial_\alpha \partial_\alpha \phi$ dominates the left hand side of \ScalarPerturb, one can combine \StressCL\ and \ScalarPerturb\ to obtain
 \eqn{AlmostNewDM}{
  \ddot\delta_p + {2\dot{a} \over a} \dot\delta_p &\approx
   \rho \sum_q {d\log m_p \over d\phi} {d\log m_q \over d\phi}
    f_q \delta_q  \cr
  &= {\rho \over 2 M_{\rm Pl}^2} \sum_q \beta_{pq} f_q \delta_q \,,
 }
where we have used $\beta_{pq} = 1 + 2M_{\rm Pl}^2 {d\log m_p \over d\phi} {d\log m_q \over d\phi}$ and $\sum_q f_q \delta_q = 0$.  Recalling that
 \eqn{RhoEquation}{
  {\rho \over 2M_{\rm Pl}^2} = 4\pi G \rho = 
   {3 \over 2} {\dot{a}^2 \over a^2} \,,
 }
we see that \AlmostNewDM\ agrees with \NewDM\ for the isocurvature modes.  Since we justified \NewDM\ for the adiabatic mode as well, its derivation as the late time limit for arbitrary perturbations is complete.

\newpage

\bibliographystyle{ssg}
\bibliography{scalar}

\end{document}